# MyElas: An automatized tool-kit for high-throughput calculation, post-processing and visualization of elasticity and related properties of solids

Hao Wang[1, 2], Y. C. Gan[2], Hua Y. Geng[2, 3*], Xiang-Rong Chen[1*]

[1] *College of Physics, Sichuan University, Chengdu 610065, People's Republic of China*

[2] *National Key Laboratory of Shock Wave and Detonation Physics, Institute of Fluid Physics, CAEP, Mianyang 621900, People's Republic of China*

[3] *HEDPS, Center for Applied Physics and Technology, and college of Engineering, Peking University, Beijing 100871, People's Republic of China*

**Abstract:** Elasticity is one of the most fundamental mechanical properties of solid. In high-throughput design of advanced materials, there is an imperative demand for the capability to quickly calculate and screen a massive pool of candidate structures. A fully automatized pipeline with minimal human intervention is the key to provide high efficiency to achieve the goal. Here, we introduce a tool-kit MyElas that aims to address this problem by forging all pre-processing, elastic constant and other related property calculations, and post-processing into an integrated framework that automatically performs the assigned tasks to drive data flowing through parallelized pipelines from input to output. The core of MyElas is to calculate the second and third order elastic constants of a solid with the energy-strain method from first-principles. MyElas can auto-analyze the elastic constants, to derive other related physical quantities. Furthermore, the tool-kit also integrates a visualization function, which can, for example, plot the spatial anisotropy of elastic modulus and sound velocity of monocrystalline. The validity and efficiency of the toolkit are tested and bench-marked on several typical systems.

---

**PROGRAM SUMMARY**

*Program title*: MyElas

*CPC Library link to program files*: (to be added by Technical Editor)

*Code Ocean capsule*: (to be added by Technical Editor)

*Licensing provisions*: GNU General Public License 3 (GPL)

*Programming language*: Python 3.X

*External routines/libraries*: Numpy [1], Spglib [2], Matplotlib [3]

*Nature of problem*: Through the first-principal calculation, the second-order and third-order elastic constants of solid materials are automatically calculated, and the post-processing and visualization of single crystal and polycrystalline physical properties are carried out.

*Solution method:* Firstly, the required strain structure is automatically generated through the space group of materials and the corresponding strain matrix. Secondly, the energy of the structure is calculated by first-principles calculation software such as VASP [4]. Thirdly, the relationship between energy and strain is polynomial fitted to deduce the elastic constant of the material. Fourth, automatically derive the relevant physical properties. Finally, the distribution of mechanical modulus and sound velocity in spherical space is visualized.

*Additional comments including Restrictions and Unusual features:* Many modules of the software can be embedded into other software through simple modification, such as visualization module. The software supports finite temperature calculation with electronic temperature. In addition, the software supports the derivation of corresponding elastic constants from phonon spectrum data generated by *Phonopy* [5], *Alamode* [6] and *TDEP* [7] software through long-wave limit approximation and Christoffel equation.

# 1. Introduction

For solid materials, elasticity describes the reversible strain response to external forces within the elastic limit, and is the fundament for understanding the mechanical properties [1-3], which plays a vital role in design of advanced materials. Therefore, to compute and screen elastic constants has become one of the most frequent routine-tasks in modern material science.

So far, most calculations are limited to calculate the second-order elastic constants (SOEC) of a single crystal. However, with SOEC, we can further derive other related mechanical properties straightforwardly, such as Young's modulus, shear modulus, and corresponding sound velocity in polycrystalline samples[4-6], which are more relevant for practical applications. The spatial anisotropy distribution of mechanical modulus (and sound velocity) in single crystals[7, 8] also can be obtained, as well as, other important parameters such as the Debye temperature and the lowest thermal conductivity.

On the other hand, first-principles methods are in-dispensable to calculate elastic constants accurately across a wide pressure range, in which SOEC can be obtained by fitting to the energy-strain or stress-strain relations[4, 9]. The elastic constants are then extracted from the second order derivatives of the energy function in the former case, and from the first order derivatives of the stress in the latter case, with respect to strains. At first glance, the stress-strain method seems simpler and easier than the former one. But in actual applications, the algorithm based on energy is much more robust than that of stress, mainly because the larger uncertainty in the numerical value of the commutated stress.

There are already several tools that can calculate the SOEC for solids, including *ElaStic*[10], *AELAS*[11], *ELATE*[8], *Elastic3rd*[12], etc. However, they all are for the ground state at zero kelvin, and are not equipped for finite temperatures. Furthermore, they suffer other limitations, especially in high-throughput non-intervention calculations. For example, *AELAS* can only deal with polycrystalline modulus. Many other important physical properties that are relevant in experiments or engineering applications, e.g., the sound velocity, Debye temperature and anisotropic distribution of monocrystalline modulus, are not evaluated. On the other hand, *ELATE* is only for online analysis and visualization of the anisotropy of SOEC, and it cannot calculate the elastic constants by itself. Elastic3rd devotes mainly to the computation of the third-order elastic constants (TOEC), but does not have a feature for post-processing and analysis of these quantities. For all of these tools, frequent manual interventions are often requisite.

In this article, we present an efficient and integrated elasticity computation and processing toolkit, MyElas, for both zero kelvin and finite temperature cases. It combines the computation of SOEC, TOEC and other elastic properties, the pre-processing of the input, the post-processing and visualization of the output into single package. MyElas is designed with parallelized high-throughput computation in mind, and forges these operations into automatic pipelines so that can deal with massive candidate structures simultaneously without hanuman intervention. In next section, we present the theory for SOEC and TOEC computation, as well as the derivation of other relevant physical quantities. In the third section, the implementation and workflow of the toolkit are described, as well as a brief introduction to the control parameters and an example of Si. The fourth section discusses testing and benchmarking of the toolkit, with the fifth section concludes the full text.

## 2. Theoretical methods

### 2.1. Elasticity theory

Elastic constants can be calculated with the energy-strain method. In the spatial coordinate system, if the initial coordinate of the material element is $a_j$, and the coordinate after a uniform elastic deformation is marked as $x_i$, then this material deformation can be represented by a deformation gradient of

$$F_{ij} = \frac{\partial x_i}{\partial a_j} \tag{1}$$

The associated Lagrangian strain tensor is defined as follows[2, 13],

$$\eta_{ij} = \frac{1}{2}\left(F^T F - I\right) \qquad (2)$$

where $I$ is the unit matrix.

In nonlinear elastic theory, the energy can be expressed as a function of Lagrangian strain ($\eta_{ij}$) through the Taylor expansion,

$$E = E_0 + \frac{1}{2!}V_0 \sum_{i,j,k,l=1}^{3} C_{ijkl}\eta_{ij}\eta_{kl} + \frac{1}{3!}V_0 \sum_{ijklmn=1}^{3} C_{ijklmn}\eta_{ij}\eta_{kl}\eta_{mn} + \cdots \qquad (3)$$

where $V_0$ is the given volume, and the coefficients of Taylor expansion are the elastic constants that need to be solved[2, 13-15].

$$C_{ijkl} = \frac{1}{V_0}\frac{\partial^2 E}{\partial \eta_{ij}\partial \eta_{kl}}\bigg|_{\eta=0} \qquad \text{(SOEC)} \qquad (4)$$

$$C_{ijklmn} = \frac{1}{V_0}\frac{\partial^3 E}{\partial \eta_{ij}\partial \eta_{kl}\partial \eta_{mn}}\bigg|_{\eta=0} \qquad \text{(TOEC)} \qquad (5)$$

The relation with the traditional symbols of elastic constants (e.g., $C_{ij}$ for SOEC and $C_{ijk}$ for TOEC) is given by index reduction of Voigt notation: 11→1, 22→2, 33→3, 23→4, 13→5, 12→6.

Because Lagrangian strain mixes first-order and second-order displacement gradients, it is hard to separate the first-order from the second-order strain effects. MyElas avoids this difficulty by calculating the SOEC with Euler strain ($\varepsilon$) directly. It has a form similar to Lagrangian strain[13].

$$E = E_0 + V_0 \sum_{i}^{6} \sigma_i \varepsilon_i + \frac{1}{2!}V_0 \sum_{i,j=1}^{6} C_{ij}\varepsilon_i \varepsilon_j + \cdots \qquad (6)$$

$$C_{ij} = \frac{1}{V_0}\frac{\partial^2 E}{\partial \varepsilon_i \partial \varepsilon_j}\bigg|_{\varepsilon=0} \qquad \text{(SOEC)} \qquad (7)$$

At finite temperatures, the internal energy E must be replaced by Helmholtz free energy F, and the SOEC are given by

$$C_{ij}(T) = \frac{1}{V_0}\frac{\partial^2 F(V,T)}{\partial \varepsilon_i \partial \varepsilon_j}\bigg|_{\varepsilon=0} \quad C_{ijk}(T) = \frac{1}{V_0}\frac{\partial^3 F(V,T)}{\partial \eta_i \partial \eta_j \partial \eta_k}\bigg|_{\eta=0} \qquad (8)$$

in which $F(V, T)=E(V,T=0)+F_{vib}(V,T)+F_e(V,T)$. The vibrational free energy $F_{vib}$ can be calculated by using phonon spectra or via molecular dynamics (MD) simulations, while electronic free energy $F_e$ is readily calculated from the direct integration of the products of density of state and Fermi-Dirac distribution function or via the Mermin's finite temperature DFT[16]. It is evident that to calculate SOEC at zero kelvin and at finite temperatures are similar, for the purpose of clarity, here we present only the cases for zero kelvin, even though MyElas is functional with the feature for finite temperature SOEC. It is worth mentioning that due to large noise in MD data, in MyElas the vibrational contribution to finite temperature SOEC is evaluated from the long wavelength limit of acoustic branch, while the $F_e$ contribution is directly taken into account by using Mermin's functional.

**2.2. Polycrystalline modulus and sound velocity**

Because of the random distribution of crystal orientations grains, polycrystalline material can be viewed as isotropy, for which the bulk modulus ($B$), Young's modulus ($E$), shear modulus ($G$), and Poisson ratio ($v$) are employed to describe the elastic response. These quantities can be obtained by averaging the elastic constants of single crystals. The averaging approach used in MyElas is the Voigt–Reuss–Hill (VRH) approximation method[4-6]:

$$9B_V = (C_{11} + C_{22} + C_{33}) + 2(C_{12} + C_{23} + C_{13}) \tag{9}$$

$$15G_V = (C_{11} + C_{22} + C_{33}) - (C_{12} + C_{23} + C_{13}) + 4(C_{44} + C_{55} + C_{66}) \tag{10}$$

$$B_R = \frac{1}{(S_{11} + S_{22} + S_{33}) + 2(S_{12} + S_{23} + S_{13})} \tag{11}$$

$$G_R = \frac{15}{4(S_{11} + S_{22} + S_{33}) - 4(S_{12} + S_{23} + S_{13}) + 3(S_{44} + S_{55} + S_{66})} \tag{12}$$

$$B_H = \frac{B_V + B_R}{2}, \quad G_H = \frac{G_V + G_R}{2} \tag{13}$$

$$E_{V(R,H)} = \frac{9B_{V(R,H)}G_{V(R,H)}}{3B_{V(R,H)} + G_{V(R,H)}} \tag{14}$$

$$v_{V(R,H)} = \frac{3B_{V(R,H)} - 2G_{V(R,H)}}{2(3B_{V(R,H)} + G_{V(R,H)})} \tag{15}$$

where $S_{ij}=[C_{ij}]^{-1}$ is the compliance tensor.

For 2D materials, the Young's modulus ($E$) and Poisson ratio ($v$) are usually given by

$$E_x = \frac{C_{11}C_{22} - C_{12}^2}{C_{22}}; \quad E_y = \frac{C_{11}C_{22} - C_{12}^2}{C_{11}} \tag{16}$$

$$v_x = \frac{C_{12}}{C_{22}}; \quad v_y = \frac{C_{12}}{C_{11}} \tag{17}$$

Furthermore, the longitudinal ($c_L$), transverse ($c_s$), bulk ($c_b$), and average ($c_m$) sound velocities for an isotropic and homogenous solid can be obtained as [17, 18].

$$c_L = \sqrt{\frac{B + \frac{4}{3}G}{\rho}} \tag{18}$$

$$c_s = \sqrt{\frac{G}{\rho}} \tag{19}$$

$$c_b = \sqrt{\frac{B}{\rho}} = \sqrt{c_L^2 - \frac{4}{3}c_S^2} \tag{20}$$

$$c_m = \left[\frac{1}{3}\left(\frac{1}{c_L^3} + \frac{2}{c_S^3}\right)\right]^{-\frac{1}{3}} \tag{21}$$

in which $\rho$ is the mass density and is related to the specific volume by $\rho = \frac{1}{V}$.

### 2.3. Spatial anisotropy distribution

For anisotropic solids such as single crystals, there is no isotropic quantity such as $E$, $B$, $G$, and $v$. The spatial distribution of elastic constants is anisotropic. Nordmann *et al.*[7] provided a detailed derivation of the spatial representation. It should be reminded that the elastic matrix needs to be transformed into the Fedorov form[19]:

$$C^F = \begin{bmatrix} C_{11} & C_{12} & C_{13} & \sqrt{2}C_{14} & \sqrt{2}C_{15} & \sqrt{2}C_{16} \\ & C_{22} & C_{23} & \sqrt{2}C_{24} & \sqrt{2}C_{25} & \sqrt{2}C_{26} \\ & & C_{33} & \sqrt{2}C_{34} & \sqrt{2}C_{35} & \sqrt{2}C_{36} \\ & & & 2C_{44} & 2C_{45} & 2C_{46} \\ & & & & 2C_{55} & 2C_{56} \\ sym. & & & & & 2C_{66} \end{bmatrix} \quad (22)$$

We then have the Fedorov form for the compliance tensor:

$$S_{ij}^F = \left[ C_{ij}^F \right]^{-1} \quad (23)$$

We introduce the direction vector $d$ and a general normal vector $n$ that is vertical to $d$, as shown in Fig. 1, both are expressed as a function of Eulerian angles $(\theta, \phi, \psi)$ as

$$d = \begin{bmatrix} \sin(\phi)\cos(\theta) \\ \sin(\phi)\sin(\theta) \\ \cos(\phi) \end{bmatrix} \quad n = \begin{bmatrix} -\cos(\theta)\cos(\phi)\cos(\psi)+\sin(\theta)\sin(\psi) \\ -\sin(\theta)\cos(\phi)\cos(\psi)-\cos(\theta)\sin(\psi) \\ \sin(\phi)\cos(\psi) \end{bmatrix} \quad (24)$$

Then, we can write the related Voigt vector as

$$\delta_{ij}\vec{e_i} \otimes \vec{e_j} \Leftrightarrow I_V = \begin{bmatrix} 1 & 1 & 1 & 0 & 0 & 0 \end{bmatrix}^T \quad (25)$$

$$d \otimes d \Leftrightarrow d_V = \begin{bmatrix} d_1 d_1 & d_2 d_2 & d_3 d_3 & \sqrt{2}d_2 d_3 & \sqrt{2}d_1 d_3 & \sqrt{2}d_1 d_2 \end{bmatrix}^T \quad (26)$$

$$n \otimes n \Leftrightarrow n_V = \begin{bmatrix} n_1 n_1 & n_2 n_2 & n_3 n_3 & \sqrt{2}n_2 n_3 & \sqrt{2}n_1 n_3 & \sqrt{2}n_1 n_2 \end{bmatrix}^T \quad (27)$$

$$m_V = \begin{bmatrix} \sqrt{2}d_1 n_1 & \sqrt{2}d_2 n_2 & \sqrt{2}d_3 n_3 & d_2 n_3 + n_2 d_3 & d_1 n_3 + n_1 d_3 & d_1 n_2 + n_1 d_2 \end{bmatrix}^T \quad (28)$$

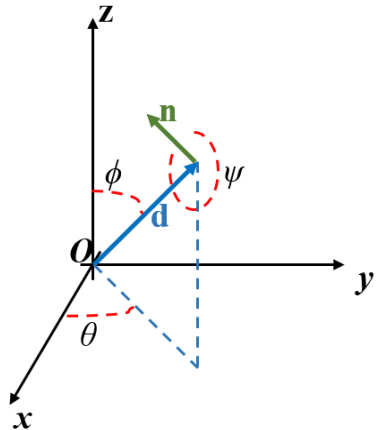

**Fig. 1.** (Color online) Spherical coordinates of a vector $d$ and normal vector $n$ in the Euclidian space.

With these Voigt vectors, the distribution of Young's modulus, bulk modulus, shear modulus, and Poisson's ratio along vector **d** and its normal vector **n** can be expressed as

$$E(d) = \left[d_V^T S^F d_V\right]^{-1} \quad (29)$$

$$B(d) = \left[3I_V^T S^F d_V\right]^{-1} \quad (30)$$

$$G(d,n) = \left[2m_V^T S^F m_V\right]^{-1} \quad (31)$$

$$v(d,n) = -E(d) d_V^T S^F d_V \quad (32)$$

Additional theoretical details can be referenced to Nordmann *et al.*[7]

The anisotropic distribution of sound velocity can be solved by using the Christoffel equation

$$\left(\Gamma_{ij} - \rho c^2 \delta_{ij}\right) u_j = 0 \quad (33)$$

$$\Gamma_{ij} = \sum_{i,j=1}^{3} C_{ijkl} d_k d_l \quad (34)$$

where $u_j$ is the displacement vector in $\vec{j}$ direction.

In a single crystal, the sound velocity obtained by Eq. (32) can be decomposed into a longitudinal mode and two orthogonal shear modes. It is often convenient to choose high symmetry directions to describe the propagation of sound. For example, for cubic system, three directions of [100], [110] and [111] are usually chosen to express the anisotropic of sound velocity. Eq. (32) is used for the derivation of the sound velocities in different lattice systems along chosen special high symmetry directions[20, 21], which are implemented in MyElas and listed below.

(1) *Cubic* crystal:

$$[100]: c_L = \sqrt{\frac{C_{11}}{\rho}}; \; c_{s1} = c_{s2} = \sqrt{\frac{C_{44}}{\rho}}$$

$$[110]: c_L = \sqrt{\frac{(C_{11}+C_{12}+2C_{44})}{2\rho}}; \; c_{s1} = \sqrt{\frac{(C_{11}-C_{12})}{2\rho}}; \; c_{s2} = \sqrt{\frac{C_{44}}{\rho}} \quad (35)$$

$$[111]: c_L = \sqrt{\frac{(C_{11}+2C_{12}+4C_{44})}{3\rho}}; \; c_{s1} = c_{s2} = \sqrt{\frac{(C_{11}-C_{12}+C_{44})}{3\rho}}$$

(2) *Hexagonal* and *Trigonal* crystal:

$$[100]: c_L = \sqrt{\frac{C_{11}-C_{12}}{2\rho}};\ c_{s1} = \sqrt{\frac{C_{11}}{\rho}};\ c_{s2} = \sqrt{\frac{C_{44}}{\rho}}$$
$$[001]: c_L = \sqrt{\frac{C_{33}}{\rho}};\ c_{s1} = c_{s2} = \sqrt{\frac{C_{44}}{\rho}}$$
(36)

(3) *Tetragonal* crystal:

$$[100]: c_L = \sqrt{\frac{C_{11}}{\rho}};\ c_{s1} = \sqrt{\frac{C_{44}}{\rho}};\ c_{s2} = \sqrt{\frac{C_{66}}{\rho}}$$
$$[001]: c_L = \sqrt{\frac{C_{33}}{\rho}};\ c_{s1} = c_{s2} = \sqrt{\frac{C_{66}}{\rho}}$$
$$[110]: c_L = \sqrt{\frac{C_{11}+C_{12}+2C_{66}}{2\rho}};\ c_{s1} = \sqrt{\frac{C_{66}}{\rho}};\ c_{s2} = \sqrt{\frac{C_{11}-C_{12}}{2\rho}}$$
(37)

(4) *Orthorhombic* crystal:

$$[100]: c_L = \sqrt{\frac{C_{11}}{\rho}};\ c_{s1} = \sqrt{\frac{C_{66}}{\rho}};\ c_{s2} = \sqrt{\frac{C_{55}}{\rho}}$$
$$[010]: c_L = \sqrt{\frac{C_{22}}{\rho}};\ c_{s1} = \sqrt{\frac{C_{66}}{\rho}};\ c_{s2} = \sqrt{\frac{C_{44}}{\rho}}$$
$$[001]: c_L = \sqrt{\frac{C_{33}}{\rho}};\ c_{s1} = \sqrt{\frac{C_{55}}{\rho}};\ c_{s2} = \sqrt{\frac{C_{44}}{\rho}}$$
(38)

### 2.4. Elasticity anisotropy index

In order to describe the anisotropy of single crystal, elastic anisotropy index has been employed in MyElas. For cubic crystal system, the anisotropy index of *Zener* ($A$) and *Chung-Buessem* ($A^c$) can be used[22].

$$A = \frac{2C_{44}}{C_{11}-C_{12}} \tag{39}$$

$$A^C = \frac{G_V - G_R}{G_V + G_R} \tag{40}$$

However, these indices are not generally applicable for all lattices. Ranganathan *et al.* proposed a universal elastic anisotropy index [23],

$$A^U = \boldsymbol{C}^V : \boldsymbol{S}^R - 6 = 5\left(\frac{G_V}{G_R} - 1\right) + \left(\frac{B_V}{B_R} - 1\right) \tag{41}$$

where $\boldsymbol{C}^V$ and $\boldsymbol{S}^R$ is the ensemble averaged stiffness and compliance tensors can be expressed in terms of the shear modulus $G$ and the bulk modulus $B$ as follows[23]:

$$\begin{aligned} \boldsymbol{C}^V &= 2G^V \boldsymbol{K} + 3B^V \boldsymbol{J} \\ \boldsymbol{S}^R &= \frac{1}{2G^R} \boldsymbol{K} + \frac{1}{3B^V} \boldsymbol{J} \end{aligned} \tag{42}$$

where $\boldsymbol{J}$ and $\boldsymbol{K}$ represent the spherical and the deviatoric parts of the unit fourth-order tensor, and the super-scripts V and R represent the Voigt and Reuss estimates, respectively. When the $A^U$ deviates from zero, the crystal anisotropy is stronger. The above anisotropy indices are a relative measure of the anisotropy with respect to a limiting value. For example, $A^U$ does not prove that a crystal having AU = 3 has double the anisotropy of another crystal with $A^U$ = 1.5 [24]. The magnitude of anisotropy between different materials cannot be determined. Therefore, Kube *et al.*[24] proposed a new scalar log-Euclidean anisotropy measure which is an absolute measure of anisotropy,

$$\begin{aligned} d_L(\boldsymbol{C}^V, \boldsymbol{C}^R) &= \|\ln \boldsymbol{C}^V - \ln \boldsymbol{C}^R\| \\ A^L(\boldsymbol{C}^V, \boldsymbol{C}^R) &= \sqrt{\left(\ln \frac{B_V}{B_R}\right)^2 + 5\left(\ln \frac{G_V}{G_R}\right)^2} \end{aligned} \tag{43}$$

where $\boldsymbol{C}^R = (\boldsymbol{S}^R)^{-1}$.

In geophysics, seismic wave anisotropy is an important index to illustrate the anisotropy of sound velocity (P-wave and S-wave)[25]:

$$AV = \frac{(V^{MAX} - V^{MIN}) \times 200}{(V^{MAX} + V^{MIN})} \tag{44}$$

where $V^{MAX}$ and $V^{MIN}$ are the maximum and minimum velocities.

On the other hand, Li *et al.* proposed the anisotropy indices to characterize the anisotropy of two-dimensional materials according to the anisotropy index of bulk materials[26]. Firstly, according to Voigt and Reuss averages, the $B^{VR}$ and $G^{VR}$ can be defined,

$$B_V = \frac{C_{11} + C_{22} + 2C_{12}}{4}; \quad G_V = \frac{C_{11} + C_{22} - 2C_{12} + 4C_{66}}{8} \quad (45)$$

$$B_R = \frac{1}{S_{11} + S_{22} + 2S_{12}}; \quad G_R = \frac{1}{S_{11} + S_{22} - 2S_{12} + S_{66}} \quad (46)$$

Based on this, Li *et al*. derived three anisotropy indices for plane materials as[26]

$$A^{SU} = \sqrt{\left(\frac{B_V}{B_R} - 1\right)^2 + 2\left(\frac{G_V}{G_R} - 1\right)^2} \quad (47)$$

$$A^{Ranganathan} = \left(\frac{B_V}{B_R} - 1\right) + 2\left(\frac{G_V}{G_R} - 2\right) \geq 0 \quad (48)$$

$$A^{Kube} = \sqrt{\left(\ln\frac{B_V}{B_R}\right)^2 + 2\left(\ln\frac{G_V}{G_R}\right)^2} \quad (49)$$

The $A^{SU}$ is used to the measure of distance between the normalized averaged stiffness tensors. $A^{Ranganathan}$ and $A^{Kube}$ is similar to the definition of $A^U$ and $A^L$ of bulk materials above. The more the three index deviates from zero, the stronger the anisotropy.

## 2.5. Other quantities

Using single crystal SOEC, other relevant physical quantity can be derived. The Debye temperature is approximated by

$$\Theta_D = \frac{h}{k_B}\left[\frac{3N}{4\pi}\left(\frac{N_A\rho}{M}\right)\right]^{\frac{1}{3}} c_m \quad (50)$$

where $h$ is the Plank constant, $k_B$ is the Boltzmann's constant, $N_A$ is the Avogadro's number, $N$ is the number of atoms in the cell, and $M$ is the mass of the cell, respectively. The averaged sound velocity $c_m$ can be calculated from Eq. (20).

The minimal thermal conductivity indicates the high temperature limit of the thermal conductivity of an insulator when the mean phonon free path is close to the interatomic distance. Based on the assumption that the average free path of phonons is comparable to the average nearest neighboring distance between atoms, there are currently two models to describe the minimum thermal conductivity of non-metallic material.

Clarke's model[27]:

$$k_{\min} = 0.87 k_B \sqrt[3]{\left(\frac{N\rho N_A}{M}\right)^2} \sqrt{\frac{E}{\rho}} \tag{51}$$

(1) Cahill's model[28]:

$$k_{\min} = \frac{k_B}{2.48} n^{\frac{2}{3}} (c_L + 2c_s) \tag{52}$$

Here, $n$ is the number density of atoms, and longitudinal sound velocity $c_L$ and shear sound velocity $c_s$ can be solved by Eqs. (17) and (18).

## 3. Implementation

### 3.1 MyElas package

MyElas is an open-source toolkit for non-commercial purposes. The package is developed mainly based on Python3.6. Here we will only briefly describe the main functions of this tool. More technique details are referenced the *user-manual* of the package. MyElas natively supports high-throughput calculations. The workflow of the calculation process is shown in Fig. 2.

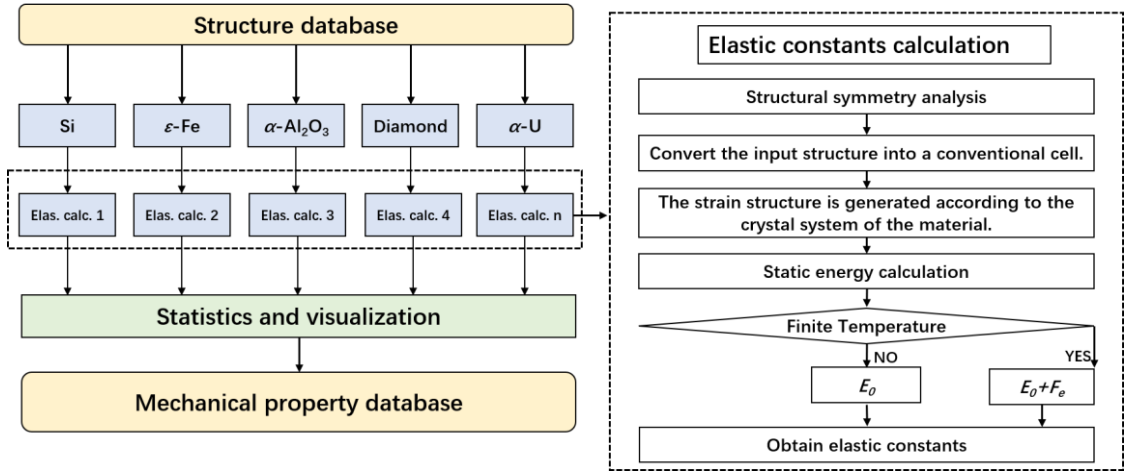

**Fig. 2.** (Color online) Work flow of MyElas in high-throughput calculation or single calculation.

MyElas provides flexible options, which are shown in Fig. 3. Through different contribution of commands, MyElas can generate required input and output files automatically.

## 3.2 Example: Si

In this subsection, we describe how to use MyElas to calculate and analyze the elastic constants and other quantities of materials, by using Si as an example. The input and output files of this example can be found in MyElas package (the folder *MyElas/examples/bulk-materials-2nd-elastic-constants/Si*).

```
usage: Myelas [-h] [-g 3D_2nd/2D_2nd/3D_3rd] [-so 3D_2nd/2D_2nd/3D_3rd] [-smax 0.018] [-snum 13]
              [-p3D Youngs/Bulk/Shear/Poisson/SV] [-p2D 2D] [-ptype 3D/plane] [-minmax min/max]
              [--vasp incar/kpoints/2D_kpoints] [--encut 600.0] [-Te 300.0] [-p 1000.0]
              [--ismear 0] [--sigma 0.05] [--isif 2]
              [-ks 0.10] [-kp 6] [-kt G/M] [-ctype rlc/stc] [-re OUTCAR/elastic.out]
              [-rp phonopy/alamode/alamode_scph/tdep]
              [-pf alamode.bands]

myelas optionn

optional arguments:
  -h, --help            show this help message and exit
  -g 3D_2nd/2D_2nd/3D_3rd, --generate 3D_2nd/2D_2nd/3D_3rd
                        Generate elastic type which need calculate
  -so 3D_2nd/2D_2nd/3D_3rd, --solve 3D_2nd/2D_2nd/3D_3rd
                        Solve elastic type which need calculate
  -smax 0.018, --strainmax 0.018
                        The strain max number
  -snum 13, --strainnum 13
                        The strain number
  -p3D Youngs/Bulk/Shear/Poisson/SV, --plot_3D Youngs/Bulk/Shear/Poisson/SV
                        plot single crystal modulus
  -p2D 2D, --plot_2D 2D
                        plot 2D materials modulus
  -ptype 3D/plane, --plot_3D_type 3D/plane
                        plot type
  -minmax min/max, --minmax min/max
                        plot maxnum or minnum for shear and poisson ratio
  --vasp incar/kpoints/2D_kpoints
                        Generate the VASP input file
  --encut 600.0         The encut energy in INCAR
  -Te 300.0, --Temperature 300.0
                        The electron temperature (K) in INCAR
  -p 1000.0, --pressure 1000.0
                        The pressure in INCAR, unit: GPa, 1GPa=10kB
  --ismear 0            The ISMEAR in INCAR
  --sigma 0.05          The smearing width in VASP
  --isif 2              The ISIF in INCAR
  -ks 0.10, --kspace 0.10
                        The KSPACING value in VASP
  -kp 6, --kpoint 6     The kpoint number in KPINTS
  -kt G/M, --kpoint_type G/M
                        The kpoint type
  -ctype rlc/stc, --calc_type rlc/stc
                        The calculation type : relax (rlx) or static (stc)
  -re OUTCAR/elastic.out, --read_elastic OUTCAR/elastic.out
                        Read the elastic tensor from OUTCAR or elastic.out file
  -rp phonopy/alamode/alamode_scph/tdep, --read_phonon_to_calc phonopy/alamode/alamode_scph/tdep
                        Read the phonon dispersiaon to calculate elastic constants
  -pf alamode.bands, --input_phonon_file alamode.bands
                        input phonon data file
```

**Fig. 3.** (Color online) The options provided in MyElas and its notes.

**Step I. Prepare the strained structures**

Firstly, we fully optimize the structure of Si to ensure that it is in the lowest energy state with command. After the optimization is completed, MyElas will rename the optimized structure file as *INPOS*. Then, MyElas is invoked again to generate the corresponding strained structures with the command:

```
Myelas -g 3D_2nd -smax 0.018 -snum 13
```

In this command, the "3D" and "0.018" indicate that the structure is bulk materials and the maximum strain value is 0.018. Three folders (*nelastic_01*, *nelastic_02*, and *nelastic_03*) will generate, corresponding to the calculation of $C_{44}$, $C_{11} + C_{12}$ and $C_{11}+2C_{12}$. Each folder contains 13 strained structures.

**Step II. Energy calculation**

When different strained structures are obtained, we need to calculate the static energy or free energy of each strained structure. At present, MyElas mainly supports VASP. In the future, we will add support for other mainstream software (e.g. *CASTEP*, *QUANTUM ESPRESSO*, *Abinit*, and *LAMMPS*). In the example of Si, the self-consistent field (SCF) convergence tolerance was set as $10^{-6}$ eV per cell (0.001 eV/Å) for energy (force), respectively. The plane-wave basis set cutoff is set as 600 eV and the K-point sampling in the irreducible Brillouin Zone (IBZ) is $2\pi\times0.1$Å$^{-1}$. The corresponding *INCAR* file for VASP can be generated on-the-fly by the following command:

```
Myelas --vasp incar –encut 600.0 -ks 0.10 -ctype stc
```

**Step III. Elastic constants extraction**

When the energy calculation is completed successfully, one can run the following command to extract the second-order elastic constants and associated physical properties of silicon,

```
Myelas -so 3D_2nd -smax 0.018 -snum 13
```

Then the output file which is called *second_elastic.out* is generated. The detailed information of this file is shown in Fig. 5. It can be seen that the obtained elastic constants of Si are well agreement with previous calculations and experimental values. In addition, in order to illustrate the fitting quality of the energy-strain curves, MyElas also outputs *nelastic.png* files for visualization (Fig. 4).

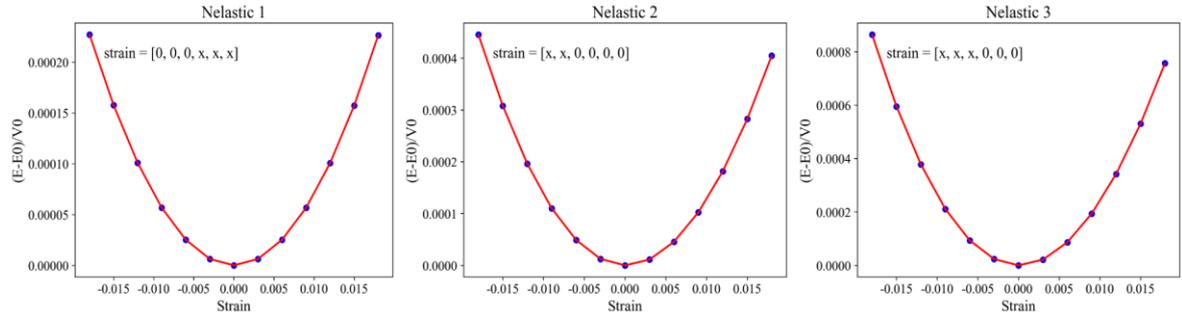

**Fig. 4.** (Color online) The strain-energy fitting plots of Si. The *Nelastic 1* (*2*, *3*) corresponds to $C_{44}$ ($C_{11} + C_{12}$ and $C_{11}+2C_{12}$), respectively. The corresponding strain tensor is also given in the figure.

```
The cubic crystal mechanical properties
Cubic crystal (spacegroup No.: 227)

Elastic tensor C_ij (unit: GPa)
  153.057   57.118   57.118    0.000    0.000    0.000
   57.118  153.057   57.118    0.000    0.000    0.000
   57.118   57.118  153.057    0.000    0.000    0.000
    0.000    0.000    0.000   74.723    0.000    0.000
    0.000    0.000    0.000    0.000   74.723    0.000
    0.000    0.000    0.000    0.000    0.000   74.723

Compliance tensor S_ij (unit: GPa^-1)
   0.008196  -0.002227  -0.002227   0.000000   0.000000   0.000000
  -0.002227   0.008196  -0.002227   0.000000   0.000000   0.000000
  -0.002227  -0.002227   0.008196   0.000000   0.000000   0.000000
   0.000000   0.000000   0.000000   0.013383   0.000000   0.000000
   0.000000   0.000000   0.000000   0.000000   0.013383   0.000000
   0.000000   0.000000   0.000000   0.000000   0.000000   0.013383

mechanical stability:  Stable

unit cell volume :   163.5297 A^3
unit cell density:   2282.2924 kg/m^3

Polycrystalline modulus
(Unit: GPa) Bulk modulus    Shear modulus    Youngs modulus    Possion ratio    P-wave modulus
  Vogit       89.0975           64.0219          154.9516          0.2101          174.4599
  Reuss       89.0975           61.0940          149.1836          0.2209          170.5561
  Hill        89.0975           62.5579          152.0676          0.2155          172.5080

Cauchy Pressure  (GPa):  -17.6056
Pugh's ratio         :   0.7021
Vickers hardness (GPa):  11.8649

Anisotropy index:
  Zener anisotropy index         :  1.56
  Chung-Buessem anisotropy index:   0.02
  Universal anisotropy index     :  0.24
  Log-Euclidean anisotropy index:  0.10

Polycrystalline sound velocity (m/s)
  Longitudinal sound velocity:   8693.9868
  Shear sound velocity       :   5235.4674
  Bulk sound velocity        :   6248.0862
  Average sound velocity     :   5789.6283

Pure single-crystal sound velocity (m/s)
  [100] direction:  vl = 8189.189   vs1 = 5721.930   vs2 = 5721.930
  [110] direction:  vl = 8876.098   vs1 = 4584.558   vs2 = 5721.930
  [111] direction:  vl = 9093.545   vs1 = 4992.555   vs2 = 4992.555

Pure single-crystal Young's modulus (GPa)
  E100 = 122.01   E010 = 122.01   E001 = 122.01   E110 = 157.98   E111 = 175.17

Debye temperature:   630.49 K

The minimum thermal conductivity (Not suitable for metallic materials):
  Clark model  :  1.312 W/(m K)
  Chaill model :  1.427 W/(m K)
```

**Fig. 5.** (Color online) The output file for calculating the second-order elastic constants of Si.

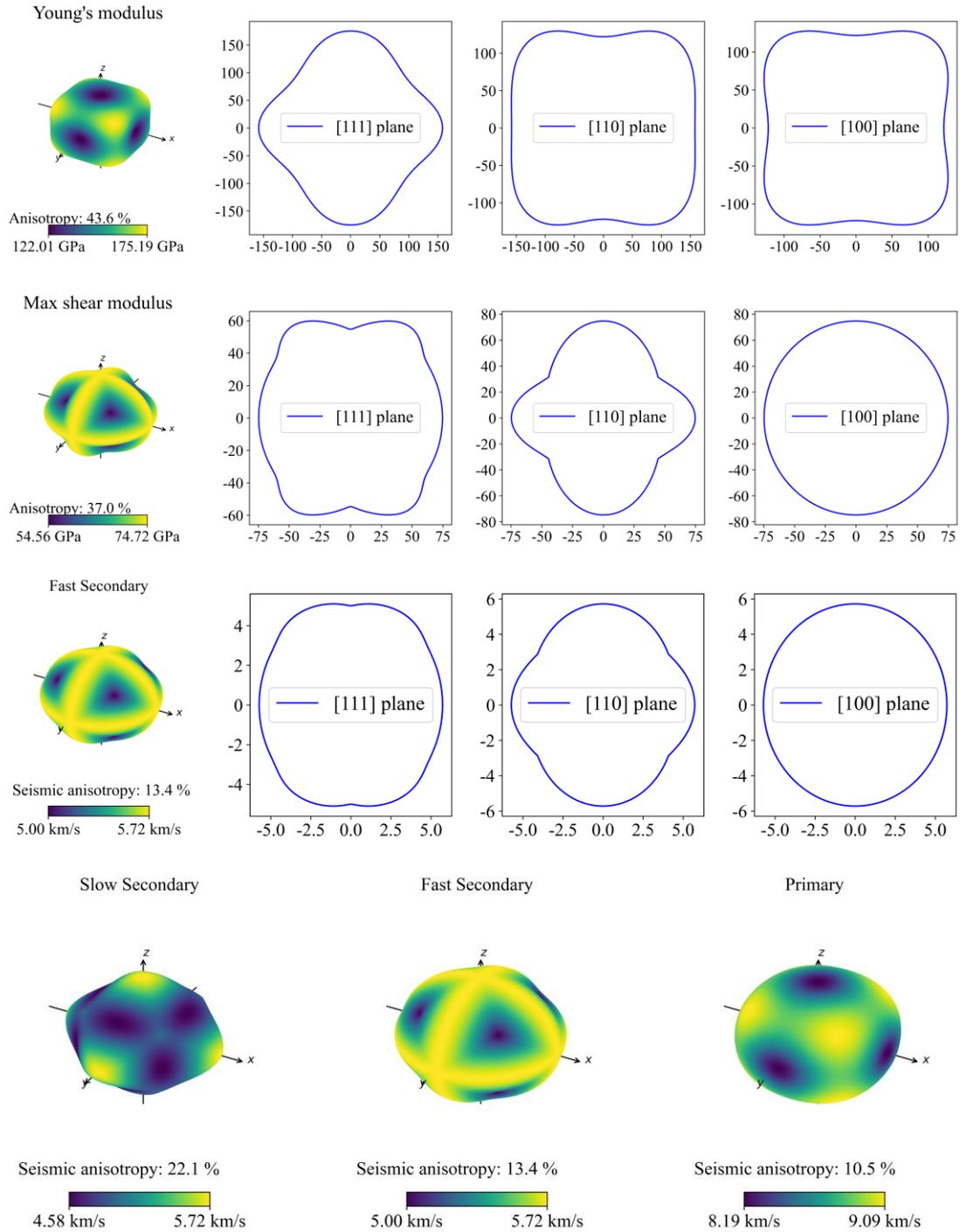

**Fig. 6.** (Color online) Visualization of Young's modulus, the maximal Shear modulus and sound velocity (include longitudinal (primary) and transverse (fast and slow secondary) sound velocity) of single-crystalline Si at 0 GPa. The projection of modulus and sound velocity onto three given planes is also shown. The colormap indicates the directional dependency of the modulus and sound velocities. The anisotropy values are also given in Figures.

**Step IV. Visualization**

In this step, in order to better analyze the anisotropy of single-crystal silicon, we can visualize its spatial distribution.

```
Myelas -p3D Youngs -ptype 3D
Myelas -p3D Shear -ptype 3D -minmax max
Myelas -p3D SV
```

Then, we will obtain the spatial anisotropic distribution of the Young's modulus, maximum shear modulus and sound velocity of single-crystal silicon (Fig. 6).

In the folder (*MyElas/script_tools*), we provide a script named *auto_elas_calculation.sh*, which can automate the above calculation process. In addition, MyElas also supports post-processing analysis of the second-order elastic constants from other software. Just prepare a simple input file (*elastic.out*) containing the elastic constants, and then run the following commands to get the corresponding processing and analysis results. See folder *MyElas/examples/read_elastic* for details. In addition, we can obtain the main elastic constants of the material from the phonon spectrum by the long wave limit approximation and Christoffel equation (Eqs. (35)-(38)). See folder *MyElas/examples/phonon_to_elas* for details.

## 4. Testing and bench-marking

### 4.1 Computational details

The testing and bench-marking calculations are performed by working with the VASP package, which is based on density functional theory[29, 30]. The electron-core interaction is described by using the projector-augmented wave (PAW) pseudopotential[31]. The electronic exchange-correlation functional is set to the generalized gradient approximation (GGA) as parameterized by Perdew, Berke, and Ernzerhof (PBE)[32]. The self-consistent field (SCF) convergence tolerance was set as $10^{-6}$ eV per cell (0.001 eV/Å) for energy (force), respectively. The plane wave basis set cutoff is set to 600 eV, and the K-point sampling in the irreducible Brillouin Zone is $2\pi \times 0.1$ Å$^{-1}$. For SOEC calculation, we uniformly use a maximum strain value of 0.018 and a strain step-size of 0.003, for calculation. where, a maximum strain value of 0.060 is used for TOEC calculation to ensure the convergence.

## 4.2 Results
### 4.2.1. The second-order elastic constants

We first compare the difference of second-order elastic constants based on Lagrangian strain and Euler strain. As shown in Table I, the elastic constants obtained from the two strains are consistent. The calculated second-order elastic constants based on the Euler strain for some typical materials are listed in Table II-VI. These results calculated by MyElas are in good agreement with the theoretical value in literature. For $\varepsilon$-Fe, all calculated data have a large difference from the experimental results, especially $C_{11}$ and $C_{33}$. In addition, there are also noticeable differences between the DFT calculation results and the experimental values of $\alpha$-U. Figure 7 plots the relative error between the calculated and the experimental data of these materials. All DFT calculations tend to overestimate the elastic constants of uranium. For $C_{55}$ the error is larger than 70%, while the errors for other elastic constants are generally between 20% ~ 40%. This deviation is believed relating to the strong electronic correlation in Fe and U, for which current DFT cannot accurately describe. These comparisons show that the calculation algorithm of second-order elastic constants of MyElas is reliable.

**Table I**. The second-order elastic constants of Si and ε-Fe with the Lagrangian strain and Euler strain. Unit: GPa.

| Materials | $C_{11}$ | $C_{33}$ | $C_{12}$ | $C_{13}$ | $C_{44}$ | |
|---|---|---|---|---|---|---|
| Si | 153.06 | | 57.12 | | 74.72 | Euler strain |
| | 153.37 | | 56.67 | | 76.35 | Lagrangian strain |
| ε-Fe | 875.8 | 977.2 | 369.8 | 317.0 | 246.6 | Euler strain |
| | 890.7 | 988.3 | 362.1 | 307.5 | 252.0 | Lagrangian strain |

**Table II**. The second-order elastic constants of Si, Diamond and MgO with the cubic structure. The unit is GPa.

| Materials | $C_{11}$ | $C_{12}$ | $C_{44}$ | Ref. |
|---|---|---|---|---|
| Si | 153.06 | 57.12 | 74.72 | This work |
|  | 153 | 57 | 75 | Calc.[33] |
|  | 166 | 64 | 80 | Expt.[34] |
| Diamond | 1055.51 | 128.64 | 562.77 | This work |
|  | 1054 | 124 | 559 | Calc.[35] |
|  | 1082 | 125 | 579 | Expt.[36] |
| MgO | 277.88 | 91.44 | 143.8 | This work |
|  | 273.41 | 90.54 | 141.38 | Calc.[37] |
|  | 297.08 | 95.36 | 156.13 | Expt.[38] |

**Table III**. The second-order elastic constants of GaN and $\varepsilon$-Fe with the hexagonal structure. The unit is GPa.

| Materials | $C_{11}$ | $C_{33}$ | $C_{12}$ | $C_{13}$ | $C_{44}$ | $C_{66}$ | Ref. |
|---|---|---|---|---|---|---|---|
| GaN | 321.51 | 356.11 | 111.91 | 79.02 | 89.61 | 104.8 | This work |
|  | 328 | 354 | 114 | 95 | 86 | 100 | Calc.[39] |
|  | 329 | 357 | 109.5 | 80 | 91 | 110 | Calc.[40] |
|  | 359.7 | 391.8 | 122.4 | 104.6 | 99.6 | 114.9 | Expt.[39] |
| $\varepsilon$-Fe | 875.8 | 977.2 | 369.8 | 317.0 | 246.6 |  | This work |
|  | 912.5 | 991.6 | 336.5 | 300.6 | 253.7 |  | Calc.[41] |
|  | 576 | 539 | 307 | 324 | 237 |  | Expt.[42] |
|  | 599(33) | 403(20) | 318(22) | 318(22) | 187(40) |  | Expt.[42] |

**Table IV**. The second-order elastic constants of $α\text{-}Al_2O_3$ with trigonal structure and $CaMoO_4$ with tetragonal structure. The unit is GPa.

| Materials | $C_{11}$ | $C_{33}$ | $C_{12}$ | $C_{13}$ | $C_{14}$ | $C_{44}$ | $C_{66}$ | Ref. |
|---|---|---|---|---|---|---|---|---|
| | 454 | 456 | 150 | 109 | 20 | 133 | 152 | This work |
| $α\text{-}Al_2O_3$ | 456 | 458 | 148 | 108 | 20 | 132 | 154 | Calc.[11] |
| | 497 | 501 | 163 | 116 | 22 | 147 | 167 | Expt.[43] |
| | $C_{11}$ | $C_{33}$ | $C_{12}$ | $C_{13}$ | $C_{16}$ | $C_{44}$ | $C_{66}$ | |
| | 130.3 | 113.8 | 53.9 | 46.8 | 9.6 | 31.0 | 37.9 | This work |
| $CaMoO_4$ | 123.4 | 109.3 | 43.9 | 48.7 | 8.1 | 31.5 | 37.4 | Calc.[10] |
| | 144.7 | 126.5 | 66.4 | 46.6 | 13.4 | 36.9 | 45.1 | Expt.[44] |

**Table V**. The second-order elastic constants of $α\text{-}U$ with orthorhombic structure. The unit is GPa.

| | $C_{11}$ | $C_{22}$ | $C_{33}$ | $C_{44}$ | $C_{55}$ | $C_{66}$ | $C_{12}$ | $C_{13}$ | $C_{23}$ | Ref. |
|---|---|---|---|---|---|---|---|---|---|---|
| | 296 | 221 | 349 | 153 | 125 | 102 | 61 | 27 | 147 | This work |
| | 296 | 216 | 367 | 153 | 129 | 99 | 60 | 29 | 141 | Calc.[45] |
| $α\text{-}U$ | 295 | 215 | 347 | 143 | 130 | 102 | 68 | 25 | 149 | Calc.[46] |
| | 299 | 231 | 364 | 150 | 132 | 100 | 59 | 30 | 144 | Calc.[47] |
| | 215 | 199 | 267 | 124 | 73 | 74 | 46 | 22 | 108 | Expt.[48] |

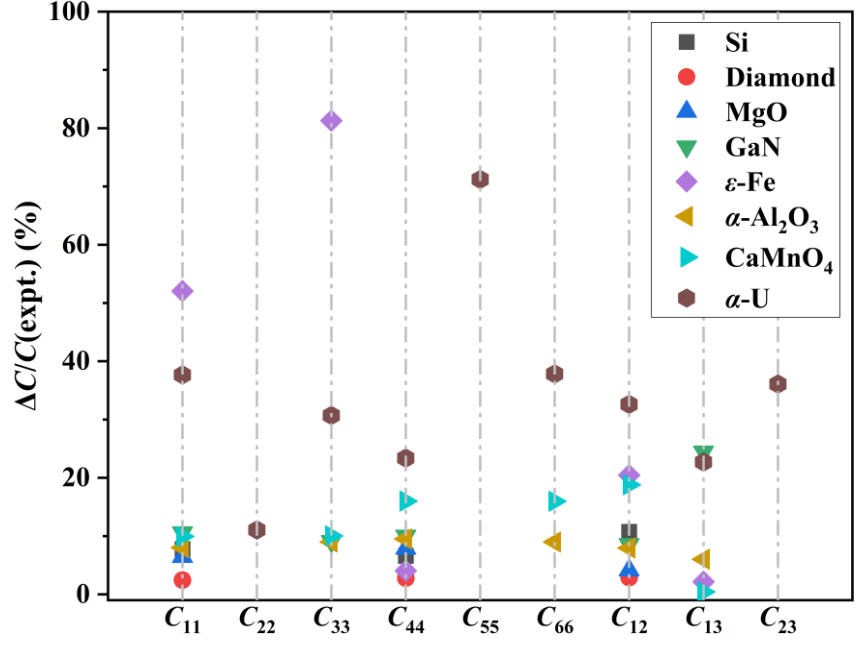

**Fig. 7.** (Color online) The relative error of elastic constants between the calculation results and experimental value.

In Table VI, the calculated elastic constants of graphene are listed, which is a 2D material. The results of MyElas are in good agreement with other theoretical data [3, 49]. For Young's modulus, our result is consistent with the experimental value 340 N/m [50].

**Table VI**. The second-order elastic constants and elastic modulus of 2D graphene. The unit is N/m, except the Poisson ratio ($v$) that is dimensionless.

|  | $C_{11}$ | $C_{12}$ | $C_{66}=G$ | $E$ | $v$ | Ref. |
|---|---|---|---|---|---|---|
|  | 354 | 60.918 | 146.877 | 343 | 0.172 | This work |
| Graphene | 358.1 | 60.4 | 148.5 |  |  | Calc.[49] |
|  | 360.84 | 66.33 | 148.8 | 350 |  | Calc.[51] |
|  |  |  |  | 340 |  | Expt.[50] |

### 4.2.2. The third-order elastic constants

For third-order elastic constants, current version of MyElas supports the cubic, hexagonal and rhombohedral systems (i.e., space group No. 149-230). The calculated results of the third-order elastic constants of Si, Diamond and ε-Fe by using MyElas are listed in Tables VII and VIII, respectively. All results are in line with literature data[15, 41, 52].

**Table VII**. The third-order elastic constants of Si and Diamond with a cubic structure. The unit is GPa.

| Materials | $C_{111}$ | $C_{112}$ | $C_{123}$ | $C_{144}$ | $C_{155}$ | $C_{456}$ | Ref. |
|---|---|---|---|---|---|---|---|
| | -746.6 | -438.7 | -92 | 60 | -315.4 | -71.3 | This work |
| Si | -810 | -422 | -61 | 31 | -293 | -61 | Calc.[15] |
| | -795 | -445 | -75 | 15 | -310 | -86 | Expt.[34] |
| | -6003.9 | -1689.6 | 596.2 | -262.3 | -3462.6 | -1383.1 | This work |
| | -5876 | -1593 | 618 | -197 | -2739 | -1111 | Calc.[52] |
| Diamond | -7750 | -2220 | 604 | -1780 | -2800 | -30 | Expt.[53] |
| | -7603 | -1909 | 835 | 1438 | -3938 | -2316 | Expt.[54] |

**Table VIII**. The third-order elastic constants of ε-Fe with the hexagonal structure. The unit is $10^1$ GPa.

| Materials | $-C_{111}$ | $-C_{222}$ | $-C_{333}$ | $-C_{112}$ | $-C_{113}$ | Ref. |
|---|---|---|---|---|---|---|
| | 982.1 | 874.6 | 902.7 | 143.8 | 89.8 | This work |
| | 1021 | 918.5 | 919.6 | 118.1 | 99.71 | Calc.[41] |
| ε-Fe | $-C_{123}$ | $-C_{133}$ | $-C_{144}$ | $-C_{155}$ | $-C_{344}$ | Ref. |
| | 0.63 | 203.1 | 83.3 | 136.8 | 259.5 | This work |
| | 18.95 | 201.2 | 47.01 | 135.2 | 220.1 | Calc.[41] |

### 4.2.3. Other related elastic properties

In Table IX, the polycrystalline modulus, sound velocity and Debye temperature of MgO and α-U are listed and compared. The calculated results of MyElas are in good agreement with other calculations [37, 46, 55]. The bulk modulus and shear modulus of α-U are overestimated slightly by comparison to the experimental value, probably due to residual strong correlation this material.

We also analyze the calculated anisotropy indices of these bulk materials. As shown in Fig. 8, the calculated anisotropy index of silicon (MgO and GaN) is consistent with Kube's results[24]. In addition, it can be seen from the figure that the cubic structure of Si and MgO has higher anisotropy than the hexagonal phase of iron and other materials.

**Table IX.** The Bulk, Young's, Shear modulus, Poisson ratio, sound velocity and Debye temperature of MgO and α-U.

|     | $B$(GPa) | $E$(GPa) | $G$(GPa) | $\sigma$ | $c_L$(km/s) | $c_s$(km/s) | $c_m$(km/s) | $\Theta_D$(K) | Ref. |
|---|---|---|---|---|---|---|---|---|---|
| MgO | 153.6 | 287.2 | 120.9 | 0.19 | 9.5 | 5.9 | 6.5 | 909 | This work |
|     | 151.5 |       | 118   |      |     |     |     |     | Calc.[37] |
|     | 163.2 |       | 126.7 |      | 9.6 | 5.9 |     | 927 | Calc.[55] |
|     | 163.1 |       | 130.2 |      | 9.7 | 6.0 |     | 940 | Expt.[56] |
| α-U | 146.1 | 265.6 | 111.0 | 0.20 | 3.9 | 2.4 | 2.6 | 287.0 | This work |
|     | 146.7 |       | 108.2 | 0.20 | 3.9 | 2.4 | 2.6 | 283.8 | Calc.[46] |
|     | 115   | 260.5 | 87    | 0.20 |     |     |     | 251 | Expt.[48] |

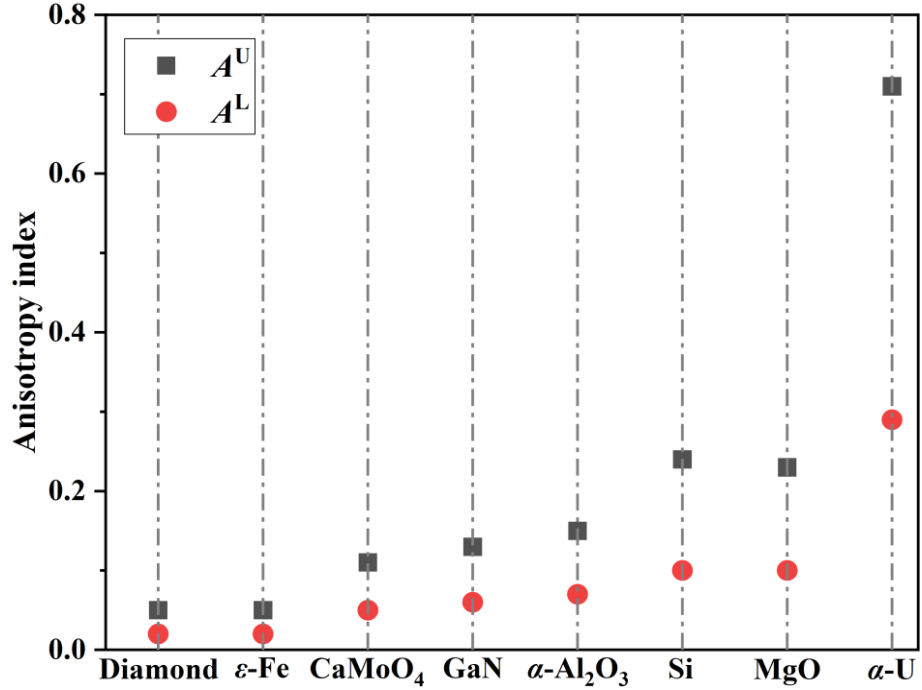

**Fig. 8.** (Color online) The anisotropy measure of different materials. Red circle indicates Log-Euclidean anisotropy index ($A^L$) and black square indicates universal anisotropy index ($A^U$).

## 5. Summary

In this work, we provide an open-source automatic elasticity calculation and analysis toolkit, MyElas. This package can not only run as an independent program, but also provide interfaces with other programs to establish parallel pipelines to form a high-throughput computing framework. Through the calculation and analysis, the results of elastic constants of different crystal systems, the reliability of the toolkit in elastic constant calculation and post-processing analysis is verified. Compared with other similar tools, MyElas can analyze and process much richer mechanical properties of materials more conveniently and quickly. This toolkit was successfully applied to predict and confirm the compression-induced softening and heating-induced hardening dual anomaly in compressed vanadium at high temperature, a validation of the finite temperature features of MyElas, which was a ready published elsewhere, and not elaborated here[57].

## Acknowledgments

This work was supported by the NSAF under Grant Nos. U1730248 and U1830101, the National Natural Science Foundation of China under Grant Nos. 11672274 and 12074214. The simulation was performed on resources provided by the Center for Comput. Mater. Sci. (CCMS) at Tohoku University, Japan.

## Competing Interests

The authors declare no competing interests.